\documentclass[prb,twocolumn,showpacs,preprintnumbers,amsmath,amssymb]{revtex4-1}

\usepackage{graphicx,rotating}
\usepackage{dcolumn}
\usepackage{bm}
\usepackage{epsfig}
\usepackage{longtable}

\begin{document}

\title{Forces and atomic relaxations in the pSIC approach with ultrasoft 
pseudopotentials.}

\author{Ma\l{}gorzata~Wierzbowska  
and Jacek~A.~Majewski}

\affiliation{
Institute of Theoretical Physics, Faculty of Physics, 
University of Warsaw, ul. Ho\.za 69, 00-681 Warszawa, Poland } 

\date{\today}

\begin{abstract}
We present the scheme that allows for efficient calculations of forces in 
the framework of pseudopotential self-interaction corrected (pSIC) 
formulation of the density functional theory. 
The scheme works with norm conserving and also with ultrasoft 
pseudopotentials and has been implemented in the plane-wave basis code 
{\sc quantum espresso}. 
We have performed tests of the internal consistency of the derived 
expressions for forces considering ZnO and CeO$_2$ crystals. 
Further, we have performed calculations of equilibrium geometry for 
LaTiO$_3$, YTiO$_3$, and LaMnO$_3$ perovskites and also for Re and Mn 
pairs in silicon. Comparison with standard DFT and DFT+U approaches 
shows that in the cases where spurious self-interaction matters, 
the pSIC approach predicts different geometry, 
very often closer to the experimental data. 
\end{abstract}

\pacs {31.15.es, 61.50.Ah, 61.72.Bb, 61.72.S-, 71.15.-m} 
%

\maketitle

\section{Introduction}

Predictive power of the density functional theory,\cite{ks} mostly in its 
local density (LDA) and gradient corrected (GGA) flavors, 
is the main factor that has established this method as the standard approach in the materials science. 
For many electronic systems, it has become possible to predict very 
accurately the equilibrium geometry, equation of state, relevant energetics, 
and further whole plethora of properties with astonishingly good accuracy. 
Unfortunately, all these approximations are plagued by the fact that 
functionals contain spurious self-interaction and the electronic states 
are typically too extended. Therefore, the reliable predictions for systems 
with very localized electronic
density, so called strongly-correlated systems, require a computational 
scheme that cures the self-interaction problem. 
%
%
In some approximate way, the self-interaction is partially removed in the
DFT+U scheme,\cite{U} which corrects the Coulomb potential within the localized
states, such as $d$- and $f$-shells of atoms. There were also developed
methods with the exact exchange,\cite{EXX-1,EXX-2,KLI} and the 
self-interaction correction (SIC).\cite{PZ,Svane,Szotek,Arai,VKP} 
Perhaps the most simple among the DFT+SIC approaches, 
is the pseudopotential SIC (pSIC) scheme proposed by 
Filippetti {\it et al.}\cite{SIC-1} 
Its usefulness to reliably predict energetics has been widely 
proved in a variety of systems, to mention just a few such as
transition metal oxides, manganites and cuprates,\cite{SIC-2} 
diluted magnetic semiconductors (DMSs),\cite{DMS,Japan} 
strongly-correlated superconductors,\cite{supercond} 
molecules\cite{molec1} and  molecular junctions,\cite{molec2} 
and many other as described in excellent review paper.\cite{SIC-2} 

Interestingly, the strongly-correlated systems exhibit very 
often strong deformations of the crystal lattice structures. 
The interesting and important examples include 
Jahn-Teller distortions, relaxations around defects, 
atomic reconstructions at interfaces, 
lattice distortions due to magnetic interactions, 
surface reconstructions and local adjustment of atomic positions 
at surfaces due to 
the adsoption of atoms and molecules, and finally clusters of atoms 
in nanoparticles.
It is obvious that the possibility to calculate forces and stress tensor, 
in addition to the energy spectrum, consistently within 
the self-interaction free DFT scheme is very desirable. 

However, unfortunately, the full equations for forces in the pSIC method 
have not been set up yet and only an attempt to calculate forces, 
albeit in a very approximate form, has been performed in the paper by Filippetti
and Fiorentini.\cite{SIC-2} Even these simplified equations for
forces have not been tested so far in any system.
Only recently, a new variational pSIC approach,\cite{new} 
different than the original pSIC approach of Filippetti and 
Spaldin,\cite{SIC-1} has been proposed.   

In this work, we provide a computational scheme that is based on 
the non-variational pSIC method,\cite{SIC-1} implementing it into widely  
used {\sc quantum espresso} code\cite{espresso} using the plane-wave basis 
and employing ultrasoft pseudopotentials (USPPs).\cite{uspp} 
For this scheme, we also derive and implement the formulae for forces. 
It turns out that the procedure to calculate forces is similar to the 
one employed in the DFT+U method.\cite{geofiz} 

The developed formalism is tested in a series of calculations for various 
systems. We calculate internal strain parameter $u$ for the wurzite ZnO and 
compare to the DFT+U results for the norm-conserving (NCPP) and 
the ultrasoft pseudopotentials. We perform tests also for the rare earth
compound CeO$_2$ with f valence shells. 
The relaxations of atomic positions in a cell are also tested for three 
chosen perovskites in distorted $P_{nma}$ structure, namely 
LaTiO$_3$, YTiO$_3$, and LaMnO$_3$.
As a third test, we consider pairs of 
Mn and Re impurities in the silicon lattice, just addressing the problem of  
transition-metal ions pairing, that is so important for a relevant 
class of materials, namely the diluted magnetic semiconductors.

The paper is organized as follows: the details of the implementation
of the pSIC are given in section II, the full equations for forces are
presented in section III, the illustrating implementations of the developed 
formalism are discussed in section IV, finally, the paper is summarized in 
section V.       

\section{Implementation of the pSIC method for plane-wave basis computational 
scheme }

In this section we describe briefly all details necessary to implement the pSIC 
scheme, just to introduce unique notation necessary for section III. 
We follow closely formulation from the work by Filippetti 
{\it et al.},\cite{SIC-1,SIC-2} and collect here the most important equations. 
Note that the second paper \cite{SIC-2} of the authors on this topic differs in some points from 
the first one, \cite{SIC-1}
mostly by setting additional simplifications which essentially 
do not affect accuracy 
but lead to a speed up of calculations. 

In the pSIC method, the Kohn-Sham equation for spin $\sigma$ orbitals 
(it implies the usage of a spin-polarized DFT approach) 
is corrected by the SIC potential $V_{SIC}$ 

\begin{equation}
[ -\nabla^{2} + \hat{V}_{PP} + \hat{V}_{HXC}^{\sigma} - \hat{V}_{SIC}^{\sigma}
] \; |\psi_{n{\bf k}}^{\sigma}\rangle = 
\varepsilon_{n{\bf k}}^{\sigma} \; |\psi_{n{\bf k}}^{\sigma}\rangle ,
\label{KS}
\end{equation}
which is cast in the Kleinman-Bylander form,\cite{KB} and 
contains contributions from the all relevant local 
orbital potentials related to the local pseudo-orbitals $\phi_i({\bf r})$ 
(with index $i$ describing lumped together angular momentum quantum numbers and position of the atom in the lattice)
as follows 

\begin{equation}
\hat{V}_{SIC}^{\sigma} = \sum_i \frac{|\gamma_i^{\sigma}\rangle
\langle \gamma_i^{\sigma}|} {C_i^{\sigma}}.
\label{Vsic}
\end{equation}

The projection operators $\gamma_i^{\sigma}$ and the normalization integrands
$C_i^{\sigma}$ are defined 

\begin{eqnarray}
\gamma_i^{\sigma}({\bf r}) & = & V_{HXC}^{\sigma}[ n_i^{\sigma} ( {\bf r})]
\; \phi_i({\bf r}) , \nonumber \\
C_i^{\sigma} & = & \langle \phi_i| \;
V_{HXC}^{\sigma}[n_i^{\sigma}] \; |\phi_i \rangle . \nonumber
\label{gamma}
\end{eqnarray}

The $V_{HXC}^{\sigma}$ potential is a sum of the Hartree potential and 
the exchange-correlation potential in a form 
that results from the DFT functional used in the calculations. 
The $V_{HXC}^{\sigma}$ potential is a functional of the local density 
$n_i^{\sigma}({\bf r})$ that is defined through  
the atomic pseudo-orbitals
$\phi_i({\bf r})$ and the occupation numbers $p_i^{\sigma}$

\begin{eqnarray}
n_i^{\sigma}({\bf r}) & =  & p_i^{\sigma} \; |\phi_i({\bf r})|^{2} , \\
p_i^{\sigma} & = &  \sum_{n{\bf k}} f_{n{\bf k}}^{\sigma} \; \langle 
\psi_{n{\bf k}}^{\sigma} | \phi_i\rangle \langle \phi_i| 
\psi_{n{\bf k}}^{\sigma} \rangle .
\label{pi}
\end{eqnarray}

The occupation numbers $p_i^{\sigma}$ are obtained like in 
the DFT+U scheme from the projection of the Kohn-Sham
states $\psi_{n{\bf k}}^{\sigma}$ onto the local atomic orbitals $\phi_i$, 
and $f_{n{\bf k}}^{\sigma}$ are 
the Fermi-Dirac occupations.  \\

Note that if the pseudo-orbital functions do not depend on spin 
(as in a spin independent PP scheme used throughout this paper), 
the spin dependence of $n_i^{\sigma}({\bf r})$ enters only via 
the occupation numbers $p_i^{\sigma}$.

It is important to perform orthonormalization of the local pseudo-orbital 
functions $\phi_i$ before using them in the above 
definition of $p_i^{\sigma}$, 
since it may change considerably the relations between the occupations
of different atomic shells.
This orthonormalization is not mandatory in the DFT+U method, since this scheme
usually involves only one shell of given atom, $d$- or $f$-shell, but 
not the both. \\

Further, the pSIC potential is scaled by one half for 
the relaxation contribution
in the extended systems\cite{SIC-1} 

\begin{equation}
V_{HXC}^{\sigma} [n_i^{\sigma}] \rightarrow \frac{1}{2} \;
V_{HXC}^{\sigma} [n_i^{\sigma}] .
\label{half}
\end{equation}

In general, the scaling coefficient is applied in this place to unify
the bulk and molecular systems.\cite{SIC-2,molec1} 

In order to simplify calculations, 
two approximations are made for the pSIC potential: \\

1) The first assumption is the linear dependence of $V_{HXC}$ 
on the occupation numbers

\begin{equation}
V_{HXC}^{\sigma} [n_i^{\sigma}]  =  p_i^{\sigma} \; V_{HXC}^{\sigma}
[n_i^{\sigma}; \; p_i^{\sigma}=1]. 
\end{equation}
Above procedure is exact for the Hartree part of the potential, 
but it is approximate for the much smaller exchange-correlation part. 
In this point the orbital exchange-correlation potential has to be 
calculated with fully spin polarized orbital density. \\

2) The second simplification assumes employing 
the spherically averaged radial local orbital density $n_i^{\sigma}(r)$ 
to compute the local orbital potential $V_{HXC}^{\sigma}$. 

Therefore, the angular part characterized by quantum number $m_l$ 
of pseudo-orbitals is used only 
to calculate $p_i^{\sigma}$ and $C^{\sigma}_{i}$ as follows 

\begin{eqnarray}
\gamma^{\sigma}_{I,m_l,l}({\bf r}) & = & \frac{1}{2} \; 
p^{\sigma}_{I,m_l,l} \; V_{HXC}^{\sigma}[n^{\sigma}_{I,l}(r);1] \;
\phi^{\sigma}_{I,m_l,l}({\bf r}),  \nonumber \\
C^{\sigma}_{I,m_l,l} & = & \frac{1}{2} \; p^{\sigma}_{I,m_l,l} 
\; \int d{\bf r} \; V_{HXC}^{\sigma}[n^{\sigma}_{I,l}(r);1]
\; (\phi^{\sigma}_{I,m_l,l}({\bf r}))^2 ,   \nonumber
\label{radial}
\end{eqnarray}
where the indices $m_l$ denote the angular momentum quantum number of 
the shell $l$ ($s$, $p$, $d$, or $f$) of the atom of type $I$. \\

The total energy within the non-variational spin polarized pSIC 
scheme is constructed to resemble the DFT one and reads 

\begin{eqnarray}
&& E_{SIC}[n,m] = 
 \sum_{i,\sigma} f_{n{\bf k}}^{\sigma} \; \varepsilon_{n{\bf k}}^{\sigma} + 
 E_{ion} - \nonumber \\
&& \sum_{\sigma} \int d{\bf r} \; n^{\sigma}({\bf r}) \;
V_{HXC}^{\sigma}[n({\bf r}), m({\bf r})] + E_{HXC}[n({\bf r}),m({\bf r})] 
- \nonumber \\ 
&& \sum_{i,\sigma}E_{HXC} [n_i^{\sigma}] + 
 \sum_{n{\bf k},\sigma}  f_{n{\bf k}}^{\sigma} \; \langle
\psi_{n{\bf k}}^{\sigma} |\hat{V}_{SIC}^{\sigma} 
| \psi_{n{\bf k}}^{\sigma} \rangle,  
\label{Etotal}
\end{eqnarray}
where $n({\bf r})$ and $m({\bf r})$ are the total and 
the spin polarization density, respectively.

The exchange-correlation part of the total energy correction is 
a small number defined as 

\begin{equation}
E_{HXC}[n_i^{\sigma}] = \int d{\bf r} \; n_i^{\sigma}({\bf r}) \left( 
\frac{1}{2} V_H[ n_i^{\sigma}({\bf r})] + \varepsilon_{XC} 
[ n_i^{\sigma}({\bf r})] \right), \nonumber 
\label{Ehxc}
\end{equation}
where $\varepsilon_{XC}$ is the local exchange-correlation energy density.

The last term in the formula~(\ref{Etotal}) is the band correction, and shifts 
the total energy very strongly, restoring its proper curvature with respect to 
a change of the lattice constant (see Fig.~7 in Ref.~[11]). \\

In the scheme presented here, we implement equations for 
the ultrasoft pseudopotentials,\cite{uspp} 
since they allow for substantial reduction of the energy cutoff 
for systems consisting of transition metals and rare earth atoms.
However, the USPP are not norm-conserving and need some additional terms to
be included in the ordinary DFT and the pSIC methods. These terms contain
the augmented charges $Q_{\alpha \alpha'}$ and 
projectors $\beta_{\alpha}$.
The overlap matrix for an orthonormality condition is 
\begin{eqnarray}
\hat{S} & = & \hat{1} + \sum_{\alpha \alpha'} 
|\beta_{\alpha}\rangle \; q_{\alpha \alpha'} \; \langle \beta_{\alpha'} |,
\label{overlap}
\end{eqnarray}
where
\begin{eqnarray}
q_{\alpha \alpha'} & = & \int d{\bf r} \; Q_{\alpha \alpha'}({\bf r}) 
\nonumber \\
Q_{\alpha \alpha'}({\bf r}) & = & \phi_{\alpha}^{AE}({\bf r})
\phi_{\alpha'}^{AE}({\bf r}) - \phi_{\alpha}^{PS}({\bf r}) 
\phi_{\alpha'}^{PS}({\bf r}) , \nonumber
\end{eqnarray} 
$\phi_{\alpha}^{AE}$ and $\phi_{\alpha}^{PS}$ are the all-electron and the
pseudo-atomic functions, and $\alpha = [n,l,m,I]$ sets all quantum numbers for 
the atom $I$. \\
 
The pseudopotential splits into the local part $V_{LOC}({\bf r})$
and the non-local part $D_{\alpha \alpha'}^{\sigma}$, which consists of
the Kleinman-Bylander term $\tilde{D}_{\alpha \alpha'}^{\sigma}$
and the augmentation term as follows 
\begin{equation}
D_{\alpha \alpha'}^{\sigma} = \tilde{D}_{\alpha \alpha'}^{\sigma} +
\int d{\bf r} \; (V_{LOC}({\bf r}) + V_{HXC}^{\sigma}({\bf r})) \;
Q_{\alpha \alpha'}({\bf r}) .  \nonumber
\end{equation}

With the above definitions, the pSIC orbital density is 

\begin{equation}
n_i^{\sigma}({\bf r})  =  p_i^{\sigma} \; ( \; 
|\phi_i({\bf r})|^{2} + 
  \sum_{\alpha \alpha'} \langle \phi_i^{\sigma} |\beta_{\alpha}\rangle
\; Q_{\alpha \alpha'}({\bf r}) \; \langle \beta_{\alpha'}
|\phi_i^{\sigma} \rangle \; ) ,  \nonumber
\end{equation}

and the pSIC-USPP occupation numbers are

\begin{eqnarray}
p_i^{\sigma} & = &  \sum_{n{\bf k}} f_{n{\bf k}}^{\sigma} \; \langle
\psi_{n{\bf k}}^{\sigma} | \phi_i\rangle \langle \phi_i|
\psi_{n{\bf k}}^{\sigma} \rangle  \times \nonumber \\
&& [ \; 1 + \sum_{\alpha \alpha'} \langle \phi_i^{\sigma} 
|\beta_{\alpha}\rangle \; q_{\alpha \alpha'} \;
\langle \beta_{\alpha'}|\phi_i^{\sigma} \rangle \; ]  \nonumber \\
& = & \sum_{n{\bf k}} f_{n{\bf k}}^{\sigma} \; \langle 
\psi_{n{\bf k}}^{\sigma} | \; \hat{S} \phi_i\rangle \langle \phi_i 
\hat{S}^{\ast} \; | \psi_{n{\bf k}}^{\sigma} \rangle .  \nonumber
\label{dens-uspp}
\end{eqnarray}

The pSIC potential within the USPP scheme contains an additional term
which reads  

\begin{eqnarray}
\hat{V}_{US}^{\sigma}  & =  & \sum_i \; \frac{1}{2} \; p_i^{\sigma} \; 
\sum_{\alpha \alpha'} \; |\beta_{\alpha} \rangle \langle \beta_{\alpha'}| 
\times \nonumber \\
&& \left( \int d{\bf r}  \; V_{HXC}^{\sigma} [n_i^{\sigma}({\bf r});1]
\; Q_{\alpha \alpha'} ({\bf r})  \; \right) \; . \nonumber
\label{VUS}
\end{eqnarray}

Thus, the Kohn-Sham equation with the USPP is  

\begin{eqnarray}
&& [-\nabla^2 + \hat{V}_{LOC} + \hat{V}_{HXC}^{\sigma} +  
\sum_{\alpha \alpha'} 
|\beta_{\alpha} \rangle D_{\alpha \alpha'}^{\sigma} \langle \beta_{\alpha'}| -
\nonumber \\
&&  (\hat{V}_{SIC}^{\sigma} + \hat{V}_{US}^{\sigma}) ] \;
| \psi_{n{\bf k}}^{\sigma} \rangle = 
\varepsilon_{n{\bf k}}^{\sigma} \; \hat{S} \; 
| \psi_{n{\bf k}}^{\sigma} \rangle .   \nonumber
\label{ks-uspp}
\end{eqnarray} \\

The total energy terms of the pSIC origin are

\begin{equation}
- \sum_{i,\sigma}E_{HXC} [n_i^{\sigma}] + 
\sum_{n{\bf k}} f_{n{\bf k}}^{\sigma} \; \langle \psi_{n{\bf k}}^{\sigma} |
 \; (\hat{V}_{SIC}^{\sigma} + \hat{V}_{US}^{\sigma})  \;
 | \psi_{n{\bf k}}^{\sigma} \rangle .  
\label{etot-uspp}
\end{equation}  

In addition, the pSIC equations in the covariant form contain the off-diagonal
occupation numbers 

\begin{eqnarray}
p^{\sigma}_{I,m_l,m'_l,l} & = & \sum_{n{\bf k}} f_{n{\bf k}} \; \langle 
\psi_{n{\bf k}}^{\sigma} | \; \hat{S} \; \phi_{I,m_l,l} \rangle
\langle \phi_{I,m'_l,l} \; \hat{S}^{\ast} \; 
| \psi_{n{\bf k}}^{\sigma} \rangle ,   \nonumber  \\
V_{SIC}^{\sigma} & = & \sum_{I,m_l,m_l',l} 
\frac{ |\gamma_{I,m_l,l} \rangle \; \frac{1}{2} \; 
p^{\sigma}_{I,m_l,m'_l,l} \; \langle \gamma_{I,m_l',l}| }
{ C_{I,m_l,l}^{1/2} \; C_{I,m_l',l}^{1/2} } 
\nonumber \\
\gamma_{I,m_l,l} & = & V_{HXC}[ n_{I,l}(r); 1] \; \phi_{I,m_l,l}({\bf r})
\nonumber \\
C_{I,m_l,l} & = & \int d{\bf r} \; \phi_{I,m_l,l}({\bf r}) \;
V_{HXC}[ n_{I,l}(r); 1] \phi_{I,m_l,l}({\bf r}). \nonumber
\label{covar}
\end{eqnarray}

Another approximation for the augmentation part of the pSIC potential 
is made, assuming that the chosen pseudo-orbitals form a complete basis set.  
Thus, the beta projetors act on the atomic radial functions and enable 
simple calculation of the radial integrals. Later, the Kohn-Sham states
 are projected onto the pseudo-orbitals in the plane-wave
representation, as it is a case in the occupation numbers.
The corresponding definitions are following

\begin{eqnarray}
\hat{V}_{US}^{\sigma}  & =  &   
\sum_{I,m_l',m_l",l}  | \; \hat{S} \; 
\phi_{I,m_l,l} \rangle \times \nonumber \\
&&  \frac{1}{2} \; p^{\sigma}_{I,m_l,m_l',l} \;
\varepsilon^{aug}_{I,m_l',m_l",l} \;
\langle \phi_{I,m_l",l} \; \hat{S}^{\ast} \; | , \nonumber 
\end{eqnarray}

and

\begin{eqnarray}
&& \varepsilon^{aug}_{I,m_l,m'_l,l}   =   
 \sum_{\alpha \alpha'} \; \langle \phi_{I,m_l,l} | 
\beta_{\alpha} \rangle \times \nonumber \\
&&  \int d{\bf r} \; V_{HXC}^{\sigma} [n_{I,l}(r);1]
\; Q_{\alpha \alpha'}({\bf r})) \;
 \langle \beta_{\alpha'} | \phi_{I,m'_l,l} \rangle . \nonumber
\label{Vaug}
\end{eqnarray}

In the above form, the $\hat{V}_{US}^{\sigma}$ potential is computationally
as simple as the occupation numbers, because the quantities 
$\varepsilon^{aug}_{i}$ depend only on the pseudopotential parameters and
can be calculated ones. 

\section{Forces in the pSIC scheme}
\label{eqsforce}

In this section, we give complete equations for forces in the pSIC scheme
with ultrasoft pseudopotentials.

According to the Hellman-Feynman theorem, the forces contain only 
the derivatives of the potentials and not the Bloch functions. 
The $\alpha$ index denotes one of the cartesian directions ${x,y,z}$ 
from now on,
and the $\alpha$ component of the displacement of atom $I$ is 
denoted as $\tau_{\alpha ,I}$. \\

Thus, following the equation (\ref{etot-uspp}), we get an expression   
for the pSIC contribution to forces 

\begin{eqnarray}
&& F_{\alpha ,I}^{SIC} \;   =   \;
- \frac{\partial E_{SIC}}{\partial \tau_{\alpha ,I}} \; = \; 
 \sum_{m_l,l,\sigma}
\frac{\partial E_{HXC}[n_{I,m_l,l}^{\sigma}]}{\partial \tau_{\alpha ,I}} 
\nonumber \\
&& \; - \; \sum_{n,{\bf k}} \; f_{n{\bf k}}^{\sigma} \;
 \left[ \left\langle \psi_{n{\bf k}}^{\sigma}
 | \frac{\partial V_{SIC}}{\partial \tau_{\alpha ,I}} |
\psi_{n{\bf k}}^{\sigma} \right\rangle \right. \; +
\left. \left\langle \psi_{n{\bf k}}^{\sigma}
 | \frac{\partial V_{US}}{\partial \tau_{\alpha ,I}} |
\psi_{n{\bf k}}^{\sigma} \right\rangle \; \right]. \nonumber
\end{eqnarray} \\

The explicit derivatives are:
\begin{eqnarray}
&&  \frac{\partial E_{HXC}[n_{I,m_l,l}^{\sigma}]}
{\partial \tau_{\alpha ,I}} \;  =  \;  
\nonumber \\
&& 2 \; \int \; d{\bf r} \; p_{I,m_l,m'_l,l}^{\sigma} \;
\frac{ \partial p_{I,m_l,m'_l,l}^{\sigma} }{\partial \tau_{\alpha ,I}} \;
[n_{I,m_l,l}({\bf r});1] \times 
\nonumber \\
&& \left(  \frac{1}{2} V_H[ n_{I,l}({\bf r});1] +  \varepsilon_{XC}
[ n_{I,l}({\bf r});1] \right)  
\label{f1}
\end{eqnarray}

and

\begin{eqnarray}
&& \frac{\partial V_{SIC}}{\partial \tau_{\alpha ,I}} \;  =
 \frac{1}{2} \; \sum_{m_l,m_l',l,\sigma}  \;\;
C_{I,m_l,l}^{-1/2} \; C_{I,m_l',l}^{-1/2}  \; \times
\nonumber \\
&&   \left[  \;
 | \gamma_{I,m_l,l} \rangle \;
\frac{ \partial p^{\sigma}_{I,m_l,m'_l,l}} {\partial \tau_{\alpha ,I}} \;
 \langle \gamma_{I,m_l',l} | \;\; + \; \right. \nonumber  \\
&& \left.
\left| \frac{\partial \gamma_{I,m_l,l}}{\partial \tau_{\alpha ,I}}
\right\rangle \; p^{\sigma}_{I,m_l,m'_l,l} \;
 \langle \gamma_{I,m_l',l}| \; + \; c.c. \;\; \right]  
\label{f2}
\end{eqnarray}

and the ultrasoft part

\begin{eqnarray}
&&  \frac{\partial V_{US}}{\partial \tau_{\alpha ,I}} \;  =  \;  
\frac{1}{2} \; \sum_{m_l',m_l",l,\sigma} \;\;
\varepsilon^{aug}_{I,m_l',m_l",l}  \;\; 
\nonumber \\
&& \left[   \; | \hat{S}\phi_{I,m_l,l} \rangle \;
\frac{ \partial p^{\sigma}_{I,m_l,m'_l,l}} {\partial \tau_{\alpha ,I}} \;
\langle \hat{S}^{\ast}\phi_{I,m_l",l}| \;\; + \; \right. 
\nonumber \\
&&  \left.
\left|  \frac{\partial (\hat{S}\phi_{I,m_l,l})} {\partial \tau_{\alpha ,I}}
\right\rangle \; p^{\sigma}_{I,m_l,m'_l,l} \;
\langle \hat{S}^{\ast}\phi_{I,m_l",l} | \; + \; c.c. \;\; \right]. 
\label{f3}
\end{eqnarray} \\

The derivatives
$\partial p^{\sigma}_{I,m_l,m'_l,l} / \partial \tau_{\alpha ,I}$
and $|\partial (\hat{S}\phi_{I,m_l,l})/\partial \tau_{\alpha ,I}\rangle$
are defined in Ref.~[21] by eqs.~(13-19), and we give them explicitely
in the appendix.
The derivative
$|\partial \gamma_{I,m_l,l}/ \partial \tau_{\alpha ,I} \rangle $
is obtained in the same way as the derivative
$|\partial (\hat{S}\phi_{I,m_l,l})/ \partial \tau_{\alpha ,I} \rangle $,
because the potential $V_{HXC}[ n_{I,l}(r)]$
moves together with the atomic functions.

For the derivative of the overlap operator $\hat{S}$, 
the following approximation is made. It is assumed
that contributions of the beta functions centred at the atoms different
than the moved atom are neglected. It turns out that this approximation
does not corrupt the accuracy, and it will be shown in the test cases later on.
This simplification is necessary, 
because in the pSIC scheme the projectors used in the
definition of the occupation numbers have to be orthogonalized,
which in turn sets a difficulty in calculation of the derivatives.

Above definitions are valid for the non-variational pSIC approach. First
approximate equations for forces have been given by Filippetti and 
Fiorentini,\cite{SIC-2} however their formalae neglected terms with 
the derivatives of the occupation numbers.
Recent work by Filippetti {\it et al.}\cite{new} 
for the variational pSIC scheme contains similar expressions for forces.
We have added the derivatives of occupation numbers in a way akin to 
the equations for forces in the DFT+U scheme.\cite{geofiz} 
These terms are rather small, and we show their effect discussing 
the CeO$_2$ case in the next section.   

\section{Tests for forces and relaxations}

\subsection{Wurzite ZnO and CeO$_2$ in the $F_{m3m}$ structure} 

As a first test case, we employ introduced scheme for forces 
to the wurzite ZnO.
We use the ultrasoft pseudopotential, the LDA exchange-correlation 
functional in the parametrization of Perdew-Zunger, 
the kinetic energy cutoff of 35 Ry, 
and the uniform Monkhorst-Pack (6,6,6) k-mesh in these calculations.

\begin{figure}[ht]
\epsfxsize=9.5cm
\includegraphics[scale=0.35,angle=0.0]{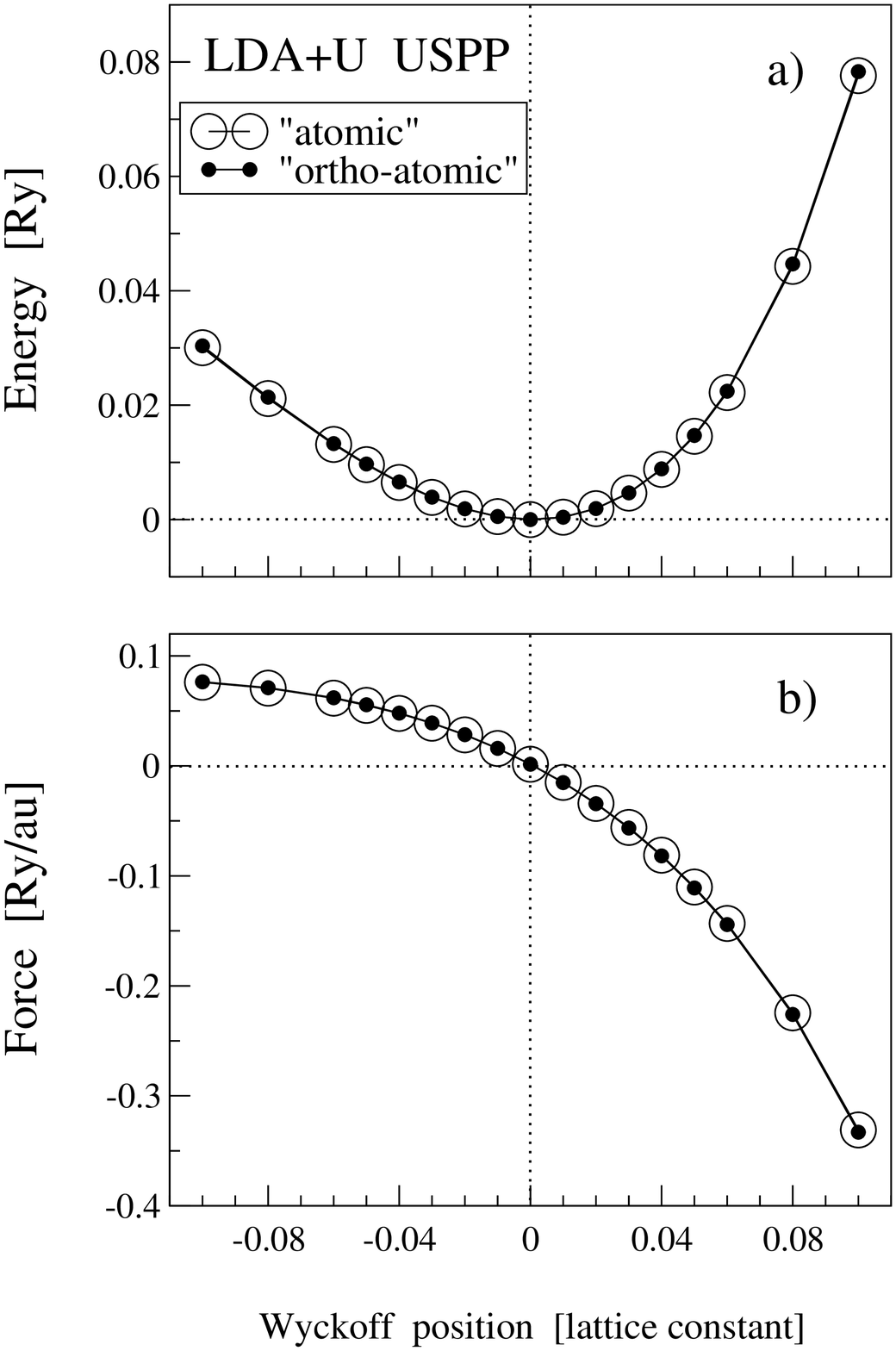}
\includegraphics[scale=0.35,angle=0.0]{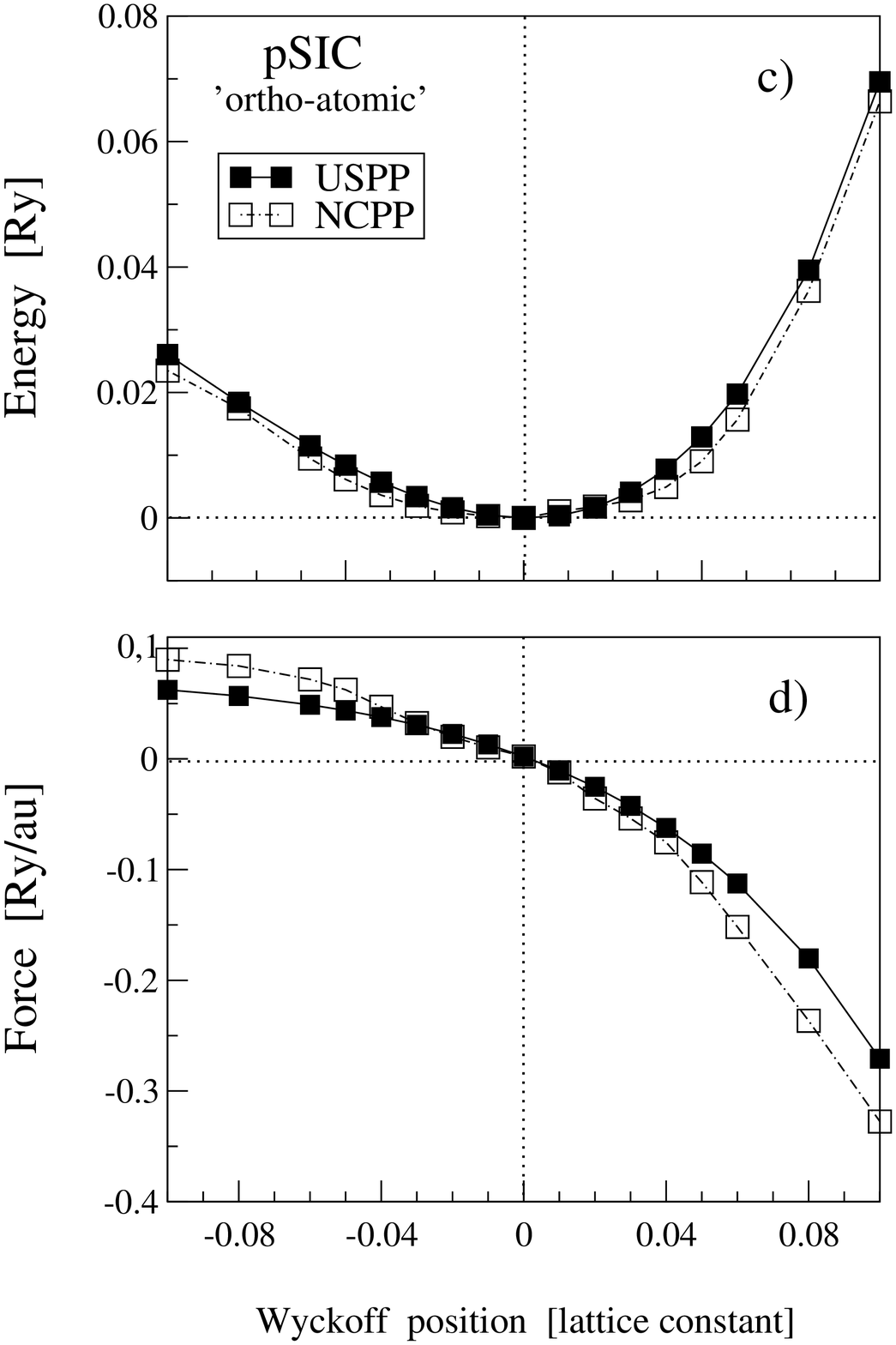}
\caption{Total energy of the wurzite ZnO and
the force acting on the displaced atom Zn(1) along z-axis.
Panels a) and b) are for the LDA+U (U=5 eV) with "atomic" and
"orthogonalized atomic" projectors for USPP,
panels c) and d) are for the LDA+pSIC with "orthogonalized atomic" projectors
for USPP and NCPP. Dotted lines are just guides for the eye.}
\label{Force}
\end{figure} 

The results for the total energy and 
the force acting on the displaced atom Zn(1) in the wurtzite unit cell 
are presented in Figure~\ref{Force}.
The Zn atom is displaced only in the $z$-direction and 
the magnitude of the displacement is given
as a function of Wyckoff position in units of the lattice constant. \\

First, we discuss the role of the approximation simplifying 
the orthogonalization 
of local atomic projectors on the total energy vs. atomic displacement curves 
and forces for both LDA+U and pSIC methods. 
As we have mentioned in the section \ref{eqsforce},
in this approximation the non-local contributions
of beta functions to the derivatives are neglected, and only the diagonal
terms in the beta functions are considered when the derivative with respect 
to the atomic position is calculated.
The LDA+U calculations (with U=5 eV) with non-orthogonalized projectors,
called "atomic", are performed without any approximation.
Simultaneously, calculations of the approximate forces obtained with
the orthogonalized projectors, called "ortho-atomic", are compared
to results from the exact formulae. Panels a) and b) of Figure~\ref{Force}
show a perfect agreement between the results for the two sets of projectors
applied for the $d$-shell,
ensuring us that the applied approximation for the derivatives in forces 
is rather good.

In panels c) and d) of Figure~\ref{Force}, the pSIC results are presented
for the same atomic displacements which have been described above 
for the LDA+U method.
As one can see, the force vanishes exactly at the geometry 
that coincides with the atomic position for which the
total energy gets the minimum. 
It clearly demonstrates the correctness of the equations for 
forces derived for the pSIC method in this paper. 
\\

Next, a relaxation of the displaced atomic positions within the wurzite
ZnO cell has been performed within the Newton-Raphson optimization scheme
based on the Broyden-Fletcher-Goldfarb-Shanno (BFGS) algorithm\cite{BFGS} 
for the estimate of the inverse Hessian matrix.
The criteria for the geometry optimization have been set as:
the energy difference between subsequent BFGS-steps $<$ 10$^{-4}$ Ry,
and the force $<$ 10$^{-3}$ Ry/a.u.

The starting non-equilibrium geometry has been 
obtained by application of the same distortion for all calculations:
the LDA, the LDA+U, and the pSIC. The wurzite structure has been perturbed in
a such way that two "$u$" parameters 
(that determine the Zn-O distances along the $z$-axis 
and are defined as the bond length along the 
hexagonal symmetry axis devised by the lattice constant $c$) 
for the Wyckoff positions have been chosen
for the Zn-O distances along the $z$-axis: 
to be equal to $u_1$=0.349 and $u_2$=0.412, which corresponds 
to considerably shorter 
and longer bond lengths, respectively (for the perfect wurzite 
structure $u_1$=$u_2$).
The lattice constant has been optimized for each method prior
to the relaxation. The identical initial distorted geometry has been used 
to find the equilibrium geometry within the LDA, LDA+U, 
and pSIC approaches. 

\begin{table}
\begin{tabular}{ccccccccc}
\hline
\hline \\
 &\;\;\;&  Exp. &\;\;\;& LDA  &\;\;\;&  LDA+U  &\;\;\;&  pSIC   \\[0.1cm]
\hline \\[-0.2cm]
lattice constant $a$  &&   6.16 &&   6.04  &&   6.05  &&    6.09   \\ 
lattice constant $c$  &&   9.84 &&   9.68  &&   9.70  &&    9.76   \\[0.1cm]
\multicolumn{7}{c}{\bf starting distorted geometry}  \\[0.1cm]
Zn(1)-O(1)          &&  \--  &&   3.382  &&   3.388   &&    3.410   \\
Zn(2)-O(2)          &&  \--  &&   3.986  &&   3.993   &&    4.019   \\[0.1cm]
\multicolumn{7}{c}{\bf relaxed parameters}  \\[0.1cm]
$u$ parameter       &&  0.382 &&  0.381   &&   0.381   &&    0.379   \\
Zn-O bond           &&  3.759 &&  3.684   &&   3.691   &&    3.697  \\[0.1cm]
\hline
\hline
\end{tabular}
\caption{The geometry parameters of initially distorted and fully 
relaxed wurzite ZnO,
calculated with the LDA, the LDA+U and the pSIC methods;
obtained with the BFGS algorithm which contains forces. Lattice constants
and bond lengths are in a.u. The experimental values are from Ref. [25].}
\label{relax}
\end{table} 

The results are displaced in Table~\ref{relax} that collects 
the optimized lattice constants, the Zn-O bond lengths along 
the $z$-axis for the distorted and relaxed structures,
and the optimized $u$ parameters. 
For all three methods, the optimized $u$ parameters (i.e., $u_1$ and $u_2$) 
are identical.
This correct result strongly corroborates 
the correctness of the derived equations for forces in the pSIC scheme. 
Note also 
that the lattice constant $c$ obtained in pSIC method agrees better 
with the experimental value (9.83 a.u.) than the lattice constants 
obtained in the LDA and the LDA+U schemes.\\

As the next test case for the equations for forces, 
we consider a rare earth compound
CeO$_2$ in the $F_{m3m}$ structure.\cite{crystal} 
Here, we have chosen for the calculations the USPP, the 
Perdew-Zunger LDA functional, the kinetic energy cutoff of 35 Ry,
and the uniform (8,8,8) Monkhorst-Pack k-mesh. 

\begin{figure}[ht]
\epsfxsize=9.5cm
\includegraphics[scale=0.32,angle=-90.0]{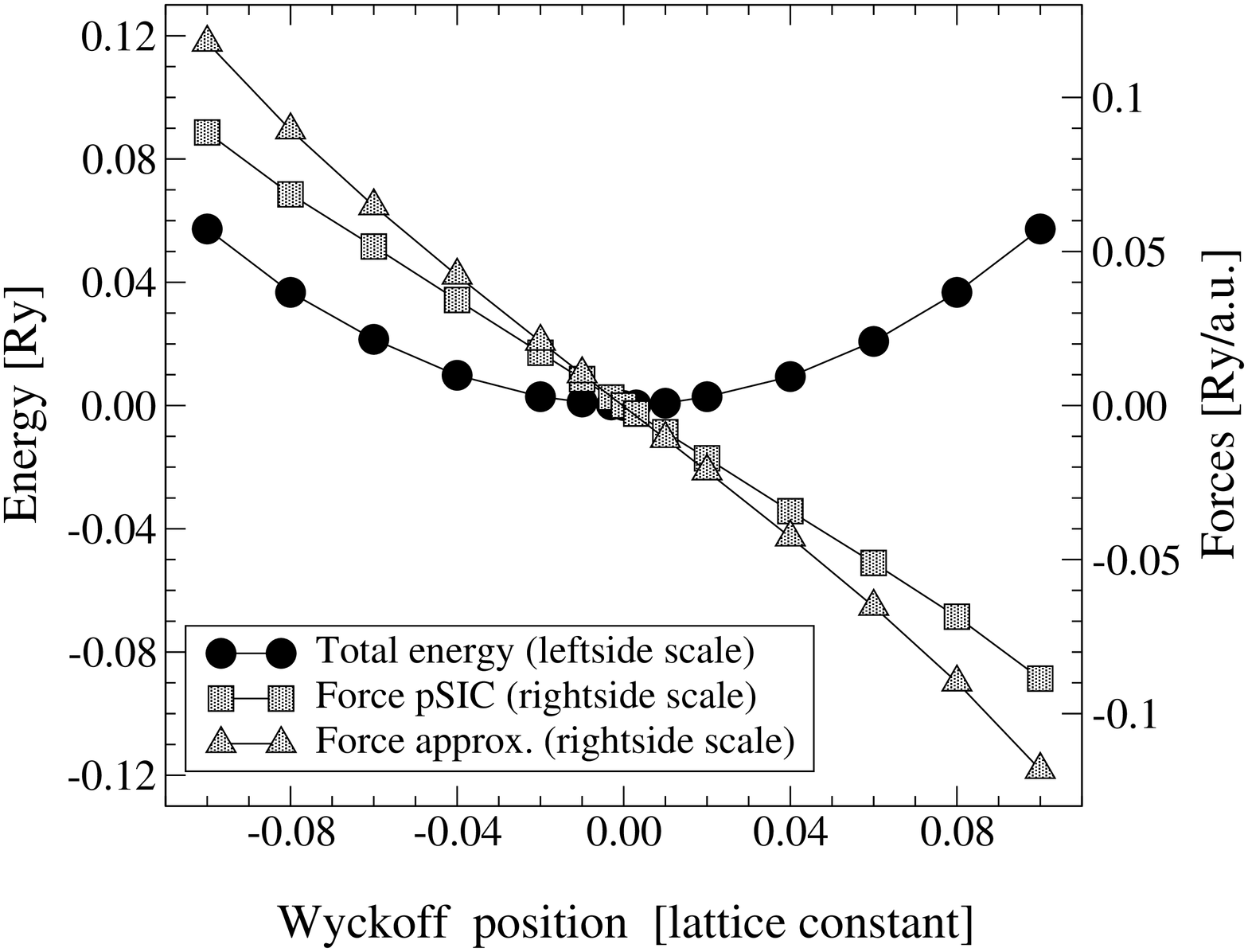}
\caption{Total energy of CeO$_2$ in the $F_{m3m}$ structure and
the forces (according to our pSIC equations -  
squares, and approximated without terms 
containing the derivatives of the occupation numbers - triangles) 
acting on the atom Ce(1) displaced along the [1,1,0] crystal-direction;
calculated with the USPP and the orthonormalized projectors.}
\label{Ce}
\end{figure}

In Figure~\ref{Ce}, the total energy and
the force acting on the displaced atom Ce(1) are shown.
The atom is displaced only along the [1,1,0] crystal-axis and the 
magnitude of the displacement is given
as a function of Wyckoff position in units of the lattice constant, which
has been fixed for this test at the experimental value of 5.41 a.u. 
The total energy minimum and the zero force occur exactly at 
the equilibrium geometry of the non-distorted structure. This
is a next proof for the derived force formulae, which work also 
for the $f$-electron 
compound. For a comparison, we have also presented approximate 
forces which have been obtained 
neglecting the derivatives with respect to the atomic position of 
the occupation numbers. Such terms enter equations
(\ref{f1}$\--$\ref{f3}), and they were omitted in Refs.~[12,18].
However, similar terms are present in the DFT+U forces.\cite{geofiz} 
In these two cases, the forces slightly differ, 
but both approaches give zero force at the same geometry. \\ 

\subsection{Distorted perovskites: LaTiO$_3$, YTiO$_3$, LaMnO$_3$}

Strongly correlated perovskites LaTiO$_3$, YTiO$_3$, and LaMnO$_3$
exhibit Jahn-Teller distorsions and crystallize in the $P_{nma}$ 
structure.\cite{crystal} 
They have been widely studied within the DFT+U method, just to 
mention as an example the work by Okatov {\it et al.}\cite{Okatov} 
(for LaTiO$_3$ and
YTiO$_3$), and by Trimarchi and Binggeli\cite{Giancarlo} (for LaMnO$_3$). 

Nevertheless, the self-interaction correction applied 
to the oxygen atom in these compounds may cause some changes in 
the predicted geometry 
in comparison to the DFT+U results. 

\begin{table}
\begin{tabular}{ccccc}
\hline
\hline \\
    Atom      &\;\;\;&     Class   &\;\;\;&   Coordinated   \\[0.1cm]
\hline \\[-0.2cm]
RE,O$_1$ && 4c && ($u$,1/4,$v$), ($\bar{u}$+1/2,3/4,$v$+1/2)   \\
         &&    && ($\bar{u}$,3/4,$\bar{v}$), ($u$+1/2,1/4,$\bar{v}$+1/2) 
 \\[0.2cm]
TM        && 4a &&  (0,0,0), (1/2,0,1/2)  \\
          &&    &&  (0,1/2,0), (1/2,1/2,1/2)  \\[0.2cm]
O$_2$     && 8d &&  $\pm$($x$,$y$,$z$) \\
          &&    &&  $\pm$($\bar{x}$,$\bar{y}$,$z$)+(1/2,0,1/2) \\
          &&    &&  $\pm$($\bar{x}$,$y$,$\bar{z}$)+(0,1/2,0) \\
          &&    &&  $\pm$($x$,$\bar{y}$,$\bar{z}$)+(1/2,1/2,1/2) \\[0.1cm]
\hline
\hline
\end{tabular}
\caption{The Wyckoff positions for each ionic specie in the $Pnma$
structure of RETMO$_3$ (RE=La,Y and TM=Ti,Mn). }
\label{perov}
\end{table}

At low temperatures, LaTiO$_3$ has a G-type antiferromagnetic structure
and YTiO$_3$ is a ferromagnet, while a colossal magnetoresistance material
LaMnO$_3$ is an A-type antiferromagnet.
It is known that relations between the cell-axes determine 
the magnetic order in distorted perovskites. 
However, the calculation of stress tensor is not implemented yet in 
the pSIC approach. 
Therefore, we focus on the FM-ordered 
structures keeping the cell parameters fixed 
at the room-temperature crystallographic data. 
The details of the $Pnma$ crystal structure are given in 
the Table~\ref{perov}. For such cell, we optimized the geometry 
employing various DFT schemes, namely the GGA, the GGA+U, and the pSIC. 

For all schemes, we have chosen the Perdew-Burke-Ernzerhof functional 
and employed the ultrasoft pseudopotentials.
In the case of the GGA+U, the Hubbard-U parameter for Ti and Mn was 
set to 3.0 eV.
In the pSIC calculations, the self-interaction correction has been applied 
to the outermost $d$-shell of rare-earth (RE) and transition-metal (TM)
elements and also to the 2$s$- and 2$p$-shells of the oxygen.
The results of calculation within the GGA,
the GGA+U, and the pSIC methods are collected in 
Table~\ref{data}, which presents crystallographic parameters obtained from 
the BFGS optimization 
and compares them to the experimental data.

\begin{table}
\begin{tabular}{crrrrrrrr}
\hline
\hline \\
 Parameters  & \;\;\; &   Exp.  & \;\;\; &  GGA  & \;\;\; & 
                          GGA+U & \;\;\; &  pSIC    \\[0.1cm]
\hline \\[-0.2cm]
\multicolumn{9}{l}{{\bf LaTiO3}} \\
\multicolumn{9}{l}{Exp.\cite{Cwik} a=10.6647 a.u., b=14.9300 a.u., 
c=10.5607 a.u.} 
\\[0.1cm]
RE $u$     &&  0.4916  &&  0.4635  &&  0.4685 &&   0.4734  \\
RE $v$     &&  0.0457  && -0.0014  && -0.0019 &&  -0.0107  \\
O$_1$ $u$  &&  0.0799  &&  0.0163  &&  0.0330 &&   0.0609  \\
O$_1$ $v$  &&  0.0087  && -0.0464  && -0.0862 &&  -0.0877  \\
O$_2$ $x$  &&  0.2096  &&  0.2022  &&  0.1938 &&   0.1997  \\
O$_2$ $y$  &&  0.0417  &&  0.0259  &&  0.0400 &&   0.0374  \\
O$_2$ $z$  &&  0.2941  &&  0.2920  &&  0.3124 &&   0.3317  \\[0.2cm]
\multicolumn{9}{l}{\bf YTiO3}  \\
\multicolumn{9}{l}{Exp.\cite{maclean} a=10.0375 a.u., b=14.3827 a.u., 
c=10.7318 a.u.} 
\\[0.1cm]
RE $u$     &&  0.4793  &&  0.4486  &&  0.4317 &&  0.4633  \\
RE $v$     &&  0.0729  && -0.0076  &&  0.0131 && -0.0109   \\
O$_1$ $u$  &&  0.1211  &&  0.0268  &&  0.0266 &&  0.0558  \\
O$_1$ $v$  &&  0.0042  && -0.0980  && -0.1227 && -0.0921   \\
O$_2$ $x$  &&  0.1910  &&  0.1852  &&  0.1864 &&  0.1791  \\
O$_2$ $y$  &&  0.0580  &&  0.0470  &&  0.0642 &&  0.0358  \\
O$_2$ $z$  &&  0.3100  &&  0.3114  &&  0.3062 &&  0.3449  \\[0.2cm]
\multicolumn{9}{l}{\bf LaMnO3}  \\
\multicolumn{9}{l}{Exp.\cite{elemans} a=10.8508 a.u., b=14.4904 a.u., 
c=10.4540 a.u.} 
\\[0.1cm]
RE $u$     &&  0.5490   &&  0.5525  &&  0.5536 &&  0.5524  \\
RE $v$     &&  0.0100   &&  0.0097  &&  0.0101 &&  0.0093  \\
O$_1$ $u$  && -0.0140   && -0.0211  && -0.0232 && -0.0255  \\
O$_1$ $v$  && -0.0700   && -0.0834  && -0.0910 && -0.0791  \\
O$_2$ $x$  &&  0.3090   &&  0.2990  &&  0.3068 &&  0.3192  \\
O$_2$ $y$  &&  0.0390   &&  0.0434  &&  0.0458 &&  0.0436  \\
O$_2$ $z$  &&  0.2240   &&  0.2144  &&  0.2180 &&  0.2274  \\[0.2cm]
\multicolumn{9}{l}{\bf Mean error $\langle \Delta \rangle$ } \\[0.1cm]  
RE $u$     &&  -  &&  0.13  &&  0.16 &&  0.08  \\
RE $v$     &&  -  &&  2.24  &&  1.65 &&  2.37  \\
O$_1$ $u$  &&  -  &&  2.08  &&  2.02 &&  1.60  \\
O$_1$ $v$  &&  -  && 30.86  && 41.42 && 34.14  \\
O$_2$ $x$  &&  -  &&  0.10  &&  0.11 &&  0.14  \\
O$_2$ $y$  &&  -  &&  0.68  &&  0.32 &&  0.60  \\
O$_2$ $z$  &&  -  &&  0.05  &&  0.10 &&  0.25  \\[0.1cm]
\hline
\hline
\end{tabular}
\caption{Experimental and theoretical parameters of $Pnma$ structure
(in crystal coordinates)
for LaTiO$_3$, YTiO$_3$ and LaMnO$_3$, obtained from the BFGS relaxation 
within the GGA, the GGA+U, and the pSIC methods. Last block in the table gives 
the mean error of parameters obtained with each method in comparison to 
the experimental values (defined by eq. (\ref{error})).}
\label{data}
\end{table}

As one can see, the distortions calculated with the pSIC method are 
usually larger than obtained from the GGA and the GGA+U methods.
Most of structural parameters calculated within the pSIC method are closer to 
the GGA+U numbers than to the GGA ones. 
Nevertheless, inclusion of the self-interaction correction
to the 2$s$ and 2$p$ shells of the oxygen leads to a substantial difference, 
and brings the pSIC results usually closer to the experimental values.
Some discrepancies still exist, especially for small parameters, 
and their reasons may lay on the accuracy of either 
the theoretical methods
or experimental techniques. On the theoretical side, for example, 
the reported calculations involve the pseudopotentials 
and it is an open question how obtained results would differ from 
the results of all-electron approach.     

Table~\ref{data} gives also 
the mean error of the calculated parameters $P^{M}$ 
with respect to the experimental values $P^{Exp}$;
it is defined as
\begin{eqnarray}
\langle \Delta \rangle & = & \sum_{struct.} 
\left|  \frac{P^M - P^{Exp} }{ P^{Exp} }  \right| ,
\label{error}
\end{eqnarray}
where the summation runs over all calculated structures: LaTiO$_3$, YTiO$_3$,
LaMnO$_3$.
One general observation is clear: the smaller is the parameter, the larger 
is the discrepancy between the calculated and the experimental values. 
Generally the distorsions from the ideal perovskite structure are larger
in the calculations than in the experiment. This might be due to the fact
that, in the experiment, the signal is averaged over the sample, which is never
clean and so ideally periodic like in the calculations.

Concerning the FM-order, all theoretical methods give 
the magnetic moments of the Ti atom equal to
 1.0 $\mu_B$ in both LaTiO$_3$ and YTiO$_3$, whereas 
the calculated magnetic moment at Mn in LaMnO$_3$ is 4.0 $\mu_B$. 
Discussion of magnetic structure issues runs beyond the scope of this work,
however, we would like to mention that the results obtained in 
this paper agree with numbers calculated within the GGA and 
the GGA+U schemes and reported earlier by other authors.\cite{Okatov} \\ 

\subsection{Diluted magnetic semiconductors: \\ Si:Mn and Si:Re}

As the third example, we have chosen two prototypes of the DMS systems. 
We consider the silicon crystal doped (i) with two Mn, 
and (ii) two Re impurities per cell. 
Detailed investigations of structural and magnetic properties of 
these DMS's will be given elsewhere. Here, we only present an 
effect of the pSIC scheme on the geometry around the 
transition-metal ions (TM) by comparing the atomic positions obtained from 
the pSIC and the standard GGA method. 
We consider two geometries of the TM pairs substituted into Si sites 
within the cubic unit cell with 64 atoms (with the silicon lattice constant 
resulting from the GGA calculations and equal to 10.32 a.u.). 
We consider (i) two TM atoms being the nearest neighbors  
(hereafter indicated as $111$, since they take the sites $(000)$ 
and $a/4(111)$ in the silicon crystal, where $a$ is the silicon
lattice constant)
and (ii) two TM atoms in the next nearest neighbours sites, 
they are bridged by the Si atom (hereafter indicated as $220$, since
they occupy the sites $(000)$ and $a/4(220)$). \\ 

\begin{table}
\begin{tabular}{ccccccccc}
\hline
\hline \\
       && \multicolumn{3}{c}{ pair 111} && 
         \multicolumn{3}{c}{ pair 220} \\[0.1cm]
 & \;\;\; & TM-TM & \; & TM-Si & \;\;\; & TM-TM & \; & TM-Si \\[0.1cm]
\hline \\[-0.2cm]   
ideal Si geom. &&  4.4686  &&  4.4686  &&  7.2983  &&  4.4686  \\[0.2cm]
GGA, TM=Mn     &&  4.8132  &&  4.5202  &&  7.1868  &&  4.4871  \\
pSIC, TM=Mn    &&  4.9495  &&  4.5429  &&  7.2735  &&  4.5552  \\[0.2cm]
GGA, TM=Re     &&  4.1837  &&  4.5532  &&  5.7524  &&  4.4768  \\ 
pSIC, TM=Re    &&  4.1713  &&  4.5181  &&  6.5140  &&  4.4108  \\[0.1cm]
\hline
\hline
\end{tabular}
\caption{The distances TM-TM and TM-Si (in Bohr) in Si:Mn and Si:Re for two  
configurations of impurities: 111 and 220 obtained after the BFGS minimization 
from the GGA and the pSIC approaches.}
\label{siremn}
\end{table}

Table~\ref{siremn} presents the distances between: (i) transition-metal ions
(TM-TM), and (ii) the transition metal and the silicon atom adjacent to 
the one of the TM-ions (for 111), and (iii) the TM-ion 
and the Si atom at the bridge TM-Si-TM (for 220); obtained from the BFGS
minimization performed in the GGA and the pSIC schemes, and compared to the
ideal geometry of the silicon crystal.

In the case of the close distance pairs (111), the Mn ions repel themselves, 
while the Re ions attract each other in comparison
to distances in the ideal silicon crystal. 
This effect is considerably stronger in the pSIC than in the GGA method. 

For the 220 pairs,
the TM ions get closer in the both cases of Mn-Mn and Re-Re pairs, 
the effect being especially pronounced for Re ions. 
In contrast to the 111 case, this attraction of TM pairs effect is 
much weaker in the pSIC than in the GGA approach.
The TM-Si distances usually become
slightly longer than the ideal Si-Si bond,
except for the Re-Si-Re bridge in the pSIC approach. 
This effect is important for the magnetic properties of silicon doped
with Re and will be published elsewhere.
Here, we only comment on the fact that, the rhenium ions in silicon have
smaller magnetic moment (1 $\mu_B$) than the Mn ions (3 $\mu_B$), and
therefore, rhenium employs more valence electrons for a hybridization 
with atoms of the host and with another close Re ion. Due to a larger 
localization of the d-shell electrons in Re within the pSIC approach, 
these states contribute much weaker to a hybridization between Re-Re,
and this bond is much longer than in the GGA method.
A very interesting difference between Si:Mn and Si:Re is 
in the DOS: the states, which are closer to the Fermi
level, originate from the closest neighbours of the impurity in the case
of Mn, and from the second close neighbours in the case of Re.
This fact gives one of the reasons why the 220 pair of Re in Si
relaxes stronger than the 111 pair.

\section{Summary}

We have derived the expressions for forces within the non-variational 
pSIC approach with ultrasoft pseudopotentials used to account for 
electron and ion interactions and implemented the scheme into the 
{\sc quantum espresso} plane-wave code. 
First, we have performed benchmark calculations to check 
the internal consistency of the scheme for the wurzite ZnO and 
rare-earth $f$-electron compound CeO$_2$ 
in the $F_{m3m}$ structure. In both cases, the forces within 
the pSIC scheme vanish for the geometry corresponding to the minimum of 
the total energy. Also optimization procedure within the code works 
perfectly bringing the initially distorted crystallographic structures 
of ZnO and CeO$_2$ into the correct equilibrium geometry efficiently.

Further, we have performed calculation within the pSIC approach to determine 
the geometry of distorted perovskites LaTiO$_3$, YTiO$_3$, and
colossal magnetoresistance compound LaMnO$_3$ in the $P_{nma}$ structure,
and also of silicon doped with pairs of Mn and Re ions. 
These systems have been chosen, since there are indications that 
the spurious self-interaction and resulting more diffused electronic 
states can lead to certain systematic errors. 
Indeed, in the cases studied here, the pSIC results for geometry parameters 
are usually 
closer to the experimental ones than the parameters obtained from 
the standard approximations of the DFT and the DFT+U methods. 
This strongly suggests that the larger localization of the electronic states 
is better accounted for in the pSIC scheme, 
which could provide also more reliable predictions in many systems. 
Also in the case of Mn and Re pairs in silicon, 
the geometries of the systems obtained within the pSIC 
and the GGA differ considerably. 
Effect of the pSIC relaxations is usually weaker than the GGA ones, which is 
a consequence of weaker $sp$-$d$ hybridization. An exception is the Re-Si-Re
configuration for which the Re1(5d)-Re2(5d) interactions are strong and
the pSIC relaxations are larger than those obtained from the GGA method. 

Having functioning scheme to calculate forces within the pSIC method, 
the further studies are under way to determine the areas of relevant 
applications and deeper investigate the reliability of the method. 

\section{Acknowledgments}
We would like to thank Andrzej Fleszar for numerous valuable discussions. 
M.W. acknowledges the support of the Leibniz Supercomputing Centre 
in Munich, where all the benchmarks have been run.
The work was supported by the European Founds for Regional Development
within the SICMAT Project (Contract No. UDA-POIG.01.03.01-14-155/09).

\begin{appendix}
\section{Derivatives of the orbital occupation numbers 
with respect to the ionic displacement} 

Partial derivatives of the occupation numbers, $p^{\sigma}_{I,m_l,m'_l,l}$,
with respect to the atomic displacements, $\tau_{\alpha ,I}$, are given
in Ref.~[21] by eqs.~(13-19). Nevertheless, for the completeness, 
we include these derivations here.  

We start from the occupation numbers in the norm-conserving pseudopotential
scheme. 
\begin{eqnarray} 
&& \frac{ \partial p^{\sigma}_{I,m_l,m'_l,l}} {\partial \tau_{\alpha ,I}} = 
\nonumber \\
&&  \sum_{n,{\bf k}} \; f_{n{\bf k}}^{\sigma} \;
 \left[  \frac{ \partial} {\partial \tau_{\alpha ,I}} 
 ( \langle \psi_{n{\bf k}}^{\sigma} | \phi_{I,m_l,l} \rangle )
 \langle \phi_{I,m_l,l} | \psi_{n{\bf k}}^{\sigma} \rangle \right. +  
\nonumber \\  
&&  \left. \langle \psi_{n{\bf k}}^{\sigma} | \phi_{I,m_l,l} \rangle
\frac{ \partial} {\partial \tau_{\alpha ,I}}
\langle \phi_{I,m_l,l} | \psi_{n{\bf k}}^{\sigma} \rangle \right] . \nonumber
\end{eqnarray}
The derivative of $\langle \psi_{n{\bf k}}^{\sigma} | \phi_{I,m_l,l} \rangle $
reduces to the derivative of $\phi_{I,m_l,l}$, since due to Hellman-Feynman
theorem $\psi_{n{\bf k}}^{\sigma}$ does not change with the displacement.

The atomic orbitals $\phi_{I,m_l,l}$ are represented in the plane-wave
basis at each vector {\bf k} from the IBZ, in order to project them onto the
Bloch functions. Then, the projection is symmetrized, to take care of the
summation over all points from the BZ. The atomic orbital at point {\bf k}
is expressed:
\begin{eqnarray}
&& \phi_{I,m_l,l,{\bf k}}({\bf r})  = \frac{1}{\sqrt{N}} 
\sum_{\bf R} e^{-i{\bf k} \cdot {\bf R}}
\phi_{I,m_l,l}({\bf r} - {\bf R} - {\bf \tau_I}) = \nonumber \\
&& e^{-i{\bf k} \cdot {\bf r}} \frac{1}{\sqrt{N}} \sum_{\bf R}
e^{-i{\bf k} ({\bf r}-{\bf R})} 
\phi_{I,m_l,l}({\bf r} - {\bf R} - {\bf \tau_I}).  \nonumber
\end{eqnarray}
N is the number of the direct lattice vectors {\bf R}. The function
$\phi_{I,m_l,l}({\bf r} - {\bf R} - {\bf \tau_I})$ is periodic with 
the lattice and its Fourier expansion in the reciprocal lattice vectors
{\bf G} is defined as:
\begin{eqnarray}
\phi_{I,m_l,l}({\bf r}) & = & \frac{1}{\sqrt{V}} 
\sum_{\bf G} e^{-i( {\bf k} + {\bf G}) \cdot {\bf r}}
c_{I,m_l,l} ({\bf k} + {\bf G}), \nonumber
\end{eqnarray}
where V is the volume of the system (V=N$\Omega$ and $\Omega$ is the cell
volume).
The Fourier components $c_{I,m_l,l} ({\bf k} + {\bf G})$ read:
\begin{eqnarray}
&& c_{I,m_l,l} ({\bf k} + {\bf G}) = 
 \frac{1}{\sqrt{V}} \int {\bf dr} \; e^{i( {\bf k} + {\bf G}) \cdot {\bf r}}
\; \phi_{I,m_l,l}({\bf r}) = \nonumber \\
&& \frac{1}{N\sqrt{\Omega}} \sum_{\bf R} \int {\bf dr} \; 
e^{i( {\bf k} + {\bf G})({\bf r}-{\bf R})} \; 
\phi_{I,m_l,l}({\bf r}- {\bf R} - {\bf \tau_I})) = \nonumber \\
&& \frac{1}{N\sqrt{\Omega}} \; e^{i( {\bf k} + {\bf G}) \cdot {\bf \tau_I}} 
\sum_{\bf R} \int {\bf dr} \; e^{i( {\bf k} + {\bf G}) \cdot {\bf r}} 
\; \phi_{I,m_l,l}({\bf r}). \nonumber
\end{eqnarray} 
The derivative of the atomic function $\phi_{I,m_l,l}$ with respect to the
displacement of the same atom I in the direction $\alpha$ is thus
\begin{eqnarray}
\frac{ \partial \phi_{I,m_l,l}}{\partial \tau_{\alpha ,I}} & = & 
 \frac{i}{\sqrt{V}} \sum_{\bf G} e^{i( {\bf k} + {\bf G}) \cdot {\bf r}}
\; c_{I,m_l,l} ({\bf k} + {\bf G}) \; ({\bf k} + {\bf G})_{\alpha}, \nonumber
\end{eqnarray} 
where $({\bf k} + {\bf G})_{\alpha}$ is the vector component along the
polarization $\alpha$.

The derivatives of the occupation numbers in the norm-conserving 
pseudopotential scheme are nonvanishing only for the displacement of the
same atom at which the occupations are considered.

In the ultrasoft-pseudopotential scheme, the derivatives
$|\partial (\hat{S}\phi_{I,m_l,l})/ \partial \tau_{\alpha ,I} \rangle $
have to be computed. According to eq. (\ref{overlap}), the above derivative 
contains derivatives of the $\beta_{\alpha}$ functions (here the index
$\alpha = [n,l,m,I]$).
These functions are the ultrasoft pseudopotential functions, which can be
expressed also in the plane-wave representation. The overlap given by
eq. (\ref{overlap}) is nonlocal in $\beta_{\alpha}$. Therefore,
we made the approximation mentioned in section III, and we neglected
contributions from the derivatives of the $\beta_{\alpha '}$ functions 
centred at atoms I' different than the moved atom I.
 
\end{appendix}

\end{document}